%% file: paper_ieee.tex
\documentclass[conference]{IEEEtran}
\IEEEoverridecommandlockouts

\input{macro_ieee.tex}
\begin{document}

\title{\tool{}: Behavioral Model Correctness Evaluation using Large Language Models
\thanks{We thank Ru Ji and Zohreh Aghababaeyan from Huawei Research Canada for their insightful discussions and valuable feedback, as well as the anonymous reviewers for their constructive comments and helpful suggestions.}
\thanks{\IEEEauthorrefmark{1}Work partially done during an internship at Huawei Research Canada.}
}

\author{\IEEEauthorblockN{Anonymous Author(s)}}

\author{\IEEEauthorblockN{Khaled Ahmed \orcidlink{0000-0001-9946-1162}}
\IEEEauthorblockA{Huawei Research Canada \\
khaled.ahmed4@h-partners.com}
\and
\IEEEauthorblockN{Jialing Song}
\IEEEauthorblockA{Huawei Research Canada \\
jackie.song@h-partners.com}
\and
\IEEEauthorblockN{Boqi Chen\IEEEauthorrefmark{1} \orcidlink{0000-0002-1451-3603}}
\IEEEauthorblockA{McGill University \\
boqi.chen@mail.mcgill.ca}
\linebreakand
\IEEEauthorblockN{Ou Wei}
\IEEEauthorblockA{Huawei Research Canada \\
ou.wei1@huawei.com}
\and
\IEEEauthorblockN{Bingzhou Zheng}
\IEEEauthorblockA{Huawei Research Canada \\
bingzhou.zheng@huawei.com}
}
\maketitle

\input{abstract}


\input{sections/intro}
\input{sections/background}
\input{sections/example}
\input{sections/approach}
\input{sections/evaluation}

\input{sections/results}
\input{sections/threats}
\input{sections/related}
\input{sections/conclusions}

\bibliographystyle{IEEEtran}
\bibliography{00}

\end{document}

%% file: macro_ieee.tex
\usepackage{cite}
\usepackage{amsmath,amssymb,amsfonts}
\usepackage{algorithmic}
\usepackage{graphicx}
\usepackage{textcomp}
\usepackage{xcolor}
\def\BibTeX{{\rm B\kern-.05em{\sc i\kern-.025em b}\kern-.08em
    T\kern-.1667em\lower.7ex\hbox{E}\kern-.125emX}}

\usepackage{soul}
\usepackage{wrapfig}
\usepackage{tikz}
\usetikzlibrary{shapes.misc, positioning}
\usepackage{multirow}
\usepackage[vskip=0.5em,font=itshape,leftmargin=1em,rightmargin=1em]{quoting}
\usepackage{framed}
\usepackage[pdfstartview=XYZ,
bookmarks=true,
colorlinks=true,
linkcolor=blue,
urlcolor=blue,
citecolor=blue,
pdftex,
bookmarks=true,
linktocpage=true, 
hyperindex=true
]{hyperref}

\usepackage{orcidlink}

\newcommand{\code}[1]{\texttt{#1}}
\newcommand{\tool}{\textsc{MC}e\textsc{T}}
\newcommand{\toolA}{\textsc{MC}e\textsc{T-A}}
\newcommand{\toolX}{\textsc{MC}e\textsc{T-X}}
\newcommand{\bench}{\textsc{FBench}}

\newcommand{\gpt}{\code{GPT-4o}}
\newcommand{\gptmini}{\code{GPT-4o-mini}}
\newcommand{\deepv}{\code{DeepSeek-v3}}
\newcommand{\deepr}{\code{DeepSeek-R1}}

\newcommand{\circled}[1]{\tikz[baseline=(char.base)]{\node[shape=circle,draw,inner sep=0.5pt,align=center] (char) {#1};}}

\newcommand{\relation}[3]{\tikz[baseline=(char.base)]{
    \node[draw,rounded rectangle] (A) {#1};
    \node (B) [right=0.15cm of A] {#2};
    \node[draw,rounded rectangle] (C) [right=0.15cm of B] {#3};

    \draw[-]
        (A) edge (B);

    \draw[->]
        (B) edge (C);
}}

\newcommand{\returnRelation}[3]{\tikz[baseline=(char.base)]{
    \node[draw,rounded rectangle] (A) {#1};
    \node (B) [right=0.15cm of A] {#2};
    \node[draw,rounded rectangle] (C) [right=0.15cm of B] {#3};

    \draw[-]
        (C) edge (B);

    \draw[->]
        (B) edge (A);
}}

\newcommand{\selfmessage}[3][text]{
    \begin{center}
    \resizebox{0.5\columnwidth}{!}{%
    \begin{tikzpicture}

        \node[] at (0,0.7) {#2};

        \draw (0,0) -- (0,0.5);
        \draw (0.1,0) -- (0.1,0.5);

        \draw[-] (0.1,0.4) -- (0.5,0.4);
        \draw[-] (0.5,0.4) -- (0.5,0.2);
        \draw[->] (0.5,0.2) -- (0.1,0.2);

        \node[right] at (0.7,0.3) {#3};
    \end{tikzpicture}
    }
    \end{center}
}

\newcommand{\selfmessagealt}[3][text]{
    \begin{center}
    \resizebox{0.6\columnwidth}{!}{%
    \begin{tikzpicture}

        \draw (-0.1,-0.1) -- (5.5,-0.1);
        \draw (5.5,-0.1) -- (5.5,0.9);
        \draw (5.5,0.9) -- (-0.1,0.9);
        \draw (-0.1,-0.1) -- (-0.1,0.9);

        \draw (0.5,0.9) -- (0.5,0.8);
        \draw (0.5,0.8) -- (0.4,0.6);
        \draw (0.4,0.6) -- (-0.1,0.6);
        \node[right] at (-0.1,0.72) {alt};

        \node[right] at (0.7,0.72) {#3};

        \draw (0,0) -- (0,0.5);
        \draw (0.1,0) -- (0.1,0.5);

        \draw[-] (0.1,0.4) -- (0.5,0.4);
        \draw[-] (0.5,0.4) -- (0.5,0.2);
        \draw[->] (0.5,0.2) -- (0.1,0.2);

        \node[right] at (0.7,0.3) {#2};

        \node[right] at (2,0) {...};
    \end{tikzpicture}
    }
    \end{center}
}

\newcommand{\messagealt}[3][text]{
    \begin{center}
    \resizebox{0.6\columnwidth}{!}{%
    \begin{tikzpicture}

        \draw (-0.1,-0.1) -- (5.7,-0.1);
        \draw (5.7,-0.1) -- (5.7,0.9);
        \draw (5.7,0.9) -- (-0.1,0.9);
        \draw (-0.1,-0.1) -- (-0.1,0.9);

        \draw (0.5,0.9) -- (0.5,0.8);
        \draw (0.5,0.8) -- (0.4,0.6);
        \draw (0.4,0.6) -- (-0.1,0.6);
        \node[right] at (-0.1,0.72) {alt};

        \node[right] at (0.7,0.72) {#3};

        \draw (0,0) -- (0,0.5);
        \draw (0.1,0) -- (0.1,0.5);

        \draw[->] (0.1,0.2) -- (4.7,0.2);

        \draw (4.7,0) -- (4.7,0.5);
        \draw (4.8,0) -- (4.8,0.5);

        \node[right] at (2,0.5) {...};

        \node[right] at (0.7,0.3) {#2};

        \node[right] at (2,0) {...};
    \end{tikzpicture}
    }
    \end{center}
}

\makeatletter
\newcommand{\linebreakand}{%
    \end{@IEEEauthorhalign}
    \hfill\mbox{}\par
    \mbox{}\hfill\begin{@IEEEauthorhalign}
}
\makeatother

\IEEEoverridecommandlockouts

\IEEEaftertitletext{\vspace{-1\baselineskip}}

%% file: abstract.tex
\begin{abstract}
Behavioral model diagrams, e.g., sequence diagrams, are an essential form of documentation that are typically designed by system engineers from requirements documentation, either fully manually or assisted by design tools.
With the growing use of Large Language Models (LLM) as AI modeling assistants, more automation will be involved in generating diagrams.
This necessitates the advancement of automatic model correctness evaluation tools.
Such a tool can be used to evaluate both manually and AI automatically generated models; to provide feedback to system engineers, and enable AI assistants to self-evaluate and self-enhance their generated models.

In this paper, we propose \tool{}, the first fully automated tool to evaluate the correctness of a behavioral model, sequence diagrams in particular, against its corresponding requirements text and produce a list of issues that the model has.
We utilize LLMs for the correctness evaluation tasks as they have shown outstanding natural language understanding ability.
However, we show that directly asking an LLM to compare a diagram to requirements finds less than 35\%  of issues that experienced engineers can find.
We propose to supplement the direct check with a fine-grained, multi-perspective approach; 
we split the diagram into \emph{atomic}, non-divisible interactions, and split the requirements text into atomic, self-contained items.
We compare the diagram with atomic requirements and each diagram-atom with the requirements.
We also propose a self-consistency checking approach that combines perspectives to mitigate LLM hallucinated issues.
Our combined approach improves upon the precision of the direct approach from 0.58 to 0.81 in a dataset of real requirements.
Moreover, the approach finds 90\% more issues that the experienced engineers found than the direct approach, and reports an average of 6 new issues per diagram.

\end{abstract}

%% file: sections/intro.tex
\section{Introduction}
\label{sec:intro}

During the process of model-driven engineering~\cite{Schmidt:MDE:Computer:2006,Kent:MDE:IFM:2002,Hutchinson:Rouncefield:Whittle:MDE:ICSE:2011},
requirement engineers iteratively design and improve the system requirements,
then develop diagrams based on these requirements.
These diagrams facilitate the communication between stakeholders, they guide and provide a structure for the implementation, and reduce the maintenance cost of the system.
Behavioral diagrams, e.g., activity diagrams, sequence diagrams, etc., are a type of diagram that detail the dynamic behavior of systems, such as the interactions between objects, state changes, and responses to external events.

Model diagrams are designed from textual requirements documents~\cite{Nuseibeh:Steve:ICSE:2000} by engineers, with or without the assistance of Artificial Intelligence (AI) tools~\cite{Ferrari:Abualhaijal:Arora:REW:2024,chen2023use,camara2023assessment,jahan2024automated}.
As both error-prone manual labor~\cite{bian2020automated,tselonis2005diagram} and hallucination-prone~\cite{Ferrari:Abualhaijal:Arora:REW:2024, Huang:Yu:Liu:Hallucination:TIS:2025, Zhang:Li:Others:Shi:arXiv:2023, Tonmoy:Zaman:Others:Das:arXiv:2024} AI agents are involved in the development of the diagrams, rigorous evaluation of their quality is needed.

Several techniques~\cite{Lilius:Paltor:ASE:1999, Latella:Majzik:Massink:FAC:1999, Jurjens:FMOODS:2002, Ledang:Souquieres:IFM:2002, Knapp:Mossakowski:CALCO:2017,Jouault:Besnard:Calvar:Teodorov:Brun:Delatour:MODELS:2020}, by transforming the model into a language with formal semantics, have been proposed to verify if the diagram satisfies behavioral properties and object constraints, or check inconsistencies with other models of the same system using model checking~\cite{Lilius:Paltor:ASE:1999, Latella:Majzik:Massink:FAC:1999}, theorem proving~\cite{Ledang:Souquieres:IFM:2002}, or consistency checks~\cite{Knapp:Mossakowski:CALCO:2017}.
However, these formal verification techniques do not evaluate the model against the textual requirements description and rely on the assumption that the diagram is already of high quality,
i.e., the diagram completely and accurately captures the textual requirements' description.
Some approaches evaluate a model against a reference written in a formal language~\cite{tselonis2005diagram, emf_diffmerge, EMFCompare, bian2019automated, hosseinibaghdadabadi2023automated, singh2022detecting, chen2024embedding, triandini2019sequence}.
Yet, these approaches fail to evaluate alternative yet valid models, and a structured formal model may not be available in many real-world scenarios.
As requirements are written in natural language and have no formal semantics, it is challenging to develop automated techniques that can evaluate the quality of the diagram against the requirements' description.

Large Language Models (LLMs) as a judge (LLM-as-a-Judge) was recently proposed for the evaluation of formal artifacts (e.g., code) against natural language~\cite{Chiang:Lee:ACL:2023, Want:Liang:Others:Zhou:2023, Zheng:Chiang:Sheng:Others:Stoica:NeurIPS:2024, Fu:Ng:Jiang:Liu:NAACL:2024} and has demonstrated reliable processing and reasoning capabilities for informal text.
We propose an automated diagram evaluation technique that evaluates the accuracy and completeness of the diagram against free-style requirements texts by utilizing the capabilities of LLMs.
However, directly evaluating the entire diagram against the requirements using LLMs does not work due to hallucinations~\cite{Huang:Yu:Liu:Hallucination:TIS:2025} and failures to focus on the entire context~\cite{Liu:Lin:Liang:LostMiddle:TACL:2024}.
We show in a preliminary study that the \textit{holistic} approach results in an inadequate evaluation,
i.e., fails to \emph{focus} on all details in the diagram~\cite{Liu:Lin:Liang:LostMiddle:TACL:2024}, and thus only reports 34\% of issues that experienced engineers are able to report, and hallucinates non-existing issues~\cite{Huang:Yu:Liu:Hallucination:TIS:2025}.

We propose a more comprehensive, fine-grained, and multi-perspective technique to evaluate diagrams: \textit{atomic checking}; we split the diagram and requirements into \emph{atomic}, indivisible interactions and short, concrete requirements, respectively.
We then compare the diagram with atomic requirements and each diagram-atom with the requirements.
Thus, the LLM can \emph{focus} on each part of the diagram and requirements, and has better chance of finding issues.
Moreover, we propose a novel self-consistency approach to mitigate hallucinations by cross-checking the results from different checks, where the hallucinated issues originate from less reliable checks are removed if they conflict with the results from more reliable ones.
This process leverages the difference of strengths between each type of check, while retaining the unique perspective of each check.
Overall, we augment the holistic check with atomic checking, and combine the results from different checks through the self-consistency approach to reduce hallucinations.

We implement our approach in a Model Correctness evaluation Tool (\tool).
We evaluate our approach on a dataset of real, industrial requirements and AI-assisted human-generated sequence diagrams, along with diagram issues reported by experienced engineers \cite{Ferrari:Abualhaijal:Arora:REW:2024}.
We show that our approach is able to detect 65\% of human-reported issues, which is 90\% more than the issues detected by the solitary holistic approach, while achieving a high precision of 81\%,
and reporting an average of 6 issues not reported by experienced engineers.

To the best of our knowledge, \tool{} is the first LLM-based approach to evaluate a behavioral diagram model against free-style requirements texts, detecting discrepancies between them, and reporting all issue explanations in natural language.
Thus, closing the automated quality assessment gap from requirements to model diagrams, helping to produce high-quality artifacts for model-driven engineering activities.

\noindent
\textbf{Contributions.} This paper makes the following contributions.

\noindent
(1) We propose the first automated behavioral model evaluation approach to evaluate a behavioral diagram model against its free-style requirements textual description.

\noindent
(2) We provide an implementation of our approach in a tool named \tool{} and make it publicly available~\cite{appendix}.

\noindent
(3) We evaluate \tool{} on a set of real requirements and sequence diagrams, and show that the tool has high precision and high human-reported issues recall.
The implementation, prompts, and the evaluation dataset used by \tool{} are available  online~\cite{appendix}.

%% file: sections/background.tex
\section{Background}
\label{sec:background}

This section describes the key concepts used in the work.

\noindent
\textbf{Requirements.}
We focus on functional requirements, which describes essential functionalities and interactions of the system.
Requirements text can be structured in several ways, some of the most common are the hierarchical fashion, i.e., starting with high-level requirements and breaking them down to sub-requirements, or grouped into use cases with preconditions, postconditions, and steps, or user stories which describe what the user wants to achieve, or in a free-style natural description.
In this work, we do not assume any structure and can process all textual requirements' descriptions.

\noindent
\textbf{Sequence diagram.}
The sequence diagram shows the interaction between the \textit{participants} of the system, e.g., users, services, and databases.
Each participant has a \textit{lifeline} representing their active period.
Participants pass around \textit{messages} in top to bottom order of the diagram, these messages carry around instructions and data.
Messages may trigger \textit{activation bars} which indicate active processing.
Messages may be shown inside \textit{combined fragments}, which represent more complex interactions, such as conditional or repeating paths.
For example, an ``alt'' block fragment shows two alternative paths, and describes the condition for each path.

\noindent
\textbf{PlantUML.}
Sequence diagrams can be designed using tools such as PlantUML~\cite{PlantUML}.
This tool provides a formal syntax for describing the diagram in a text file, and an engine that converts the syntax into a visual diagram.
Thus, automatically processing a sequence diagram designed with PlantUML is as simple as parsing the syntax, converting it to an abstract syntax tree, and performing any further processing on the tree.

\noindent
\textbf{LLM-as-a-Judge.}
Large language models (LLMs) are generative models capable of producing natural language outputs given a textual input. Typically, both input and output texts of an LLM are split into sequences of \emph{tokens}. With increased scale and complexity, LLMs have demonstrated capabilities in various \emph{reasoning} tasks, such as mathematical problem-solving, code generation, question-answering, etc.~\cite{plaat2024reasoning}.

Recently, using LLMs to evaluate the quality of inputs, known as \emph{LLM-as-a-Judge}, has gained popularity~\cite{Chiang:Lee:ACL:2023, Want:Liang:Others:Zhou:2023, Zheng:Chiang:Sheng:Others:Stoica:NeurIPS:2024, Fu:Ng:Jiang:Liu:NAACL:2024}. Such evaluation methods have been applied to various tasks, including code generation, text summarization, translation, open-domain question answering, etc.~\cite{gu2024survey,li2024llms}. However, their use in model-driven engineering remains limited.

Typically, an LLM-as-a-Judge approach generates a numeric score representing the quality of the evaluated input~\cite{lin2023llm,liu2023g}. A significant drawback of this method is its lack of explainability, as it provides limited insight into how the score was determined. To address this limitation, alternative approaches \cite{renze2024self,pan2023automatically} leverage LLMs to explicitly identify and describe specific \emph{issues} in the inputs, resulting in finer-grained and more interpretable evaluations. In this paper, we adopt this latter approach, proposing an LLM-based evaluation to detect and describe issues in sequence diagrams.



%% file: sections/example.tex
\section{\tool{} on an Example}
\label{sec:example}

\begin{figure}[htbp]
    \centering
    \includegraphics[width=\linewidth]{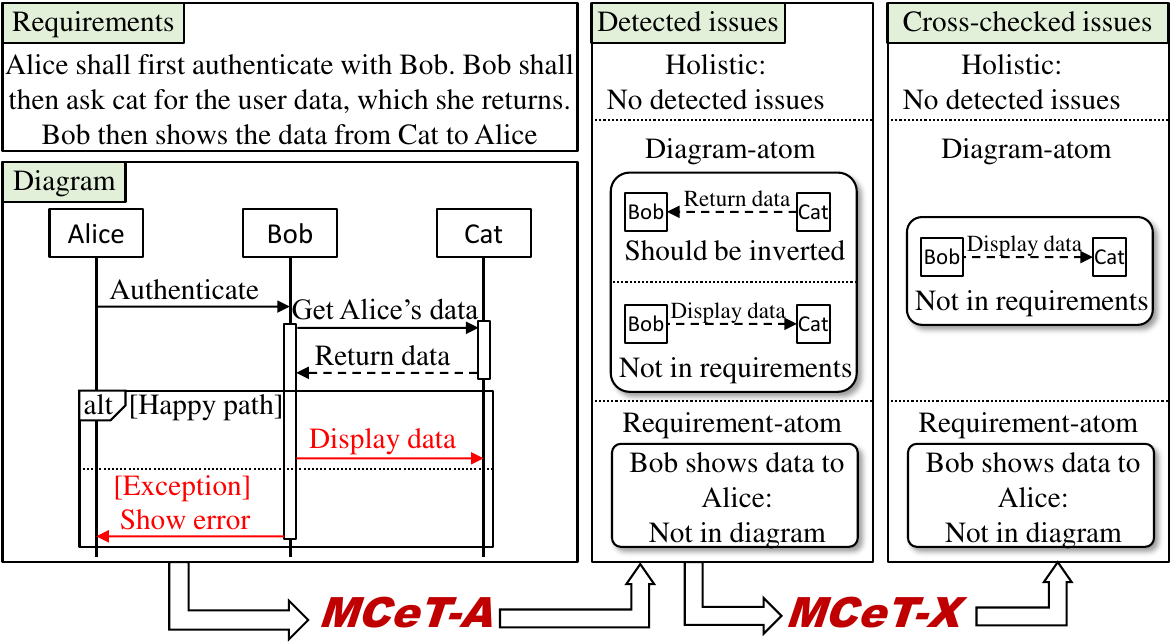}
    \caption{A requirements text, its faulty associated diagram, and the detected issues at both stages of \tool{}.}
    \label{fig:motivating_example}
    \vspace{-10pt}
\end{figure}

We explain our approach on the example in Figure~\ref{fig:motivating_example}.
The requirements describe that \textit{Alice} interacts with a system that loads and shows the user data to her.
The diagram covers most of the requirements; however, it has two problems, (1) the \textit{Display data} message from \textit{Bob} should go to \textit{Alice} instead of \textit{Cat},
and (2) the condition \textit{\lbrack Exception\rbrack} and its associated message \textit{Show error} are not part of the requirements.

Our approach aims at detecting and reporting these problems in the diagram to the user.
To this end, \tool{} uses the requirements as a reference in evaluating the quality of the diagram, but makes no assumptions about formality, structure, or style of the document, i.e., can handle free-style text, and evaluates that the diagram is both
\textit{\circled{1} Accurate}: The behavior described by the model is consistent with the requirements,
and
\textit{\circled{2} Complete}: The model covers all requirements present in the text with sufficient detail.
We borrow these definitions from prior work on manually evaluating sequence diagrams~\cite{Ferrari:Abualhaijal:Arora:REW:2024}.

A straightforward approach to evaluate this diagram is the \textit{Holistic} approach, where the entire diagram is evaluated against the requirements with one LLM invocation.
We perform a pilot study to assess the effectiveness of the holistic approach in detecting issues. The full details of this study are presented in Section~\ref{sec:evaluation} and Section~\ref{sec:results}.
The study shows that, on 16 diagrams with a total of 27 issues reported by humans, the holistic approach only detects 10 issues (37\%).
This result indicates that the holistic approach does not check every aspect of the diagram or requirements in detail.



Thus, we propose to augment the holistic approach with a more fine-grained approach;
we decompose the holistic check into its two complementary checks that it implicitly does: checking that the diagram elements accurately reflect the requirements, and checking that all requirements are completely implement by the diagram.
We split the diagram into \textit{diagram-atoms}: indivisible interactions in the diagram, which we define as a message and its two involved participants, e.g., \relation{Alice}{Authenticate}{Bob} is a diagram-atom.
Prior work has proposed several other ``atoms'' that can be checked during the manual correctness checking of sequence diagrams~\cite{Yue:Briand:Labiche:Modeling:TOSEM:2013},
e.g., incorrect messages, participants, lifelines, conditions, etc.
Our definition of a diagram-atom covers both messages and their participants, which are the main building blocks and the most frequent elements of a sequence diagram.
\tool{} then checks that the atom is accurately implemented within the diagram;
the check evaluates the accuracy of diagram's elements, i.e., the message name, its type, the participants, the direction of the interaction,
and the surrounding context of the atom which includes previous and following interactions, notes on the message, conditions, combined fragments, etc.

We also split the requirements into \textit{requirement-atoms}: indivisible concrete requirements,
which we define as a requirement that includes at most one action involving one or more participants,
e.g., \textit{Alice shall first authenticate with Bob} is a requirement-atom that involves the ``authenticate'' action only, and two participants, Alice and Bob.
\tool{} then checks that all elements and interactions mentioned in the requirement-atom are implemented in the diagram, i.e., evaluating the completeness of the diagram, and checking that the implementation is accurate by evaluating the accuracy of the related messages, participants, notes, combined fragments, and other diagram elements mentioned in the atom.

The input of \tool{} is both the requirements text and a textual format of the diagram in a modeling language.
\tool{} works in two stages, the first of which is the \textbf{A}tomic checking stage, \toolA{}, which performs three types of checks:
\textit{Holistic check}: evaluates the entire diagram against the requirements,
\textit{Diagram-atom check}: evaluates each diagram-atom in the diagram against the requirements, and
\textit{Requirement-atom check}: evaluates the diagram against each atomic requirement.

The center part of Figure~\ref{fig:motivating_example} shows an example output of the \toolA{} which is in the form of a list of issues found in the diagram, along with the localization of these issues.
For example, \toolA{} reports that the requirement-atom \textit{Bob shows data to Alice} is not implemented in the diagram,
because in the diagram, \textit{Bob} sends the \textit{Display data} to \textit{Cat} instead of Alice.
\toolA{} may, however, hallucinate some non-existing issues such as \returnRelation{Bob}{Return data}{Cat}: \textit{Should be inverted}, or miss some issues, such as \returnRelation{Alice}{Show Error}{Bob} which should not be in the diagram.

To reduce such hallucinated issues, we develop a self-consistency step that employs an authority-based \textbf{cross}-check between the results of different checks in a second stage of the tool, \toolX.
The cross-check identifies cases where a low-authority check reports an issue in the diagram but another, high-authority check marks relevant parts of the issue as correct; based on the contradiction, we deem the issue as a hallucination and remove it.
For example, the issue \returnRelation{Bob}{Return data}{Cat}: \textit{Should be inverted} was detected by the diagram-atom check, but no issues were found in the requirement-atom related to Cat returning data to Bob.
Thus, \toolX{} eliminates that issue from the final output, which is shown on the right side of Figure~\ref{fig:motivating_example}.
Next, we describe the different types of checks, discuss how the hierarchy of authority is decided for the cross-check, and show how both stages of \tool{} work.

%% file: sections/approach.tex
\section{Approach}
\label{sec:approach}

\begin{figure*}[htbp]
    \centering
    \includegraphics[width=\textwidth]{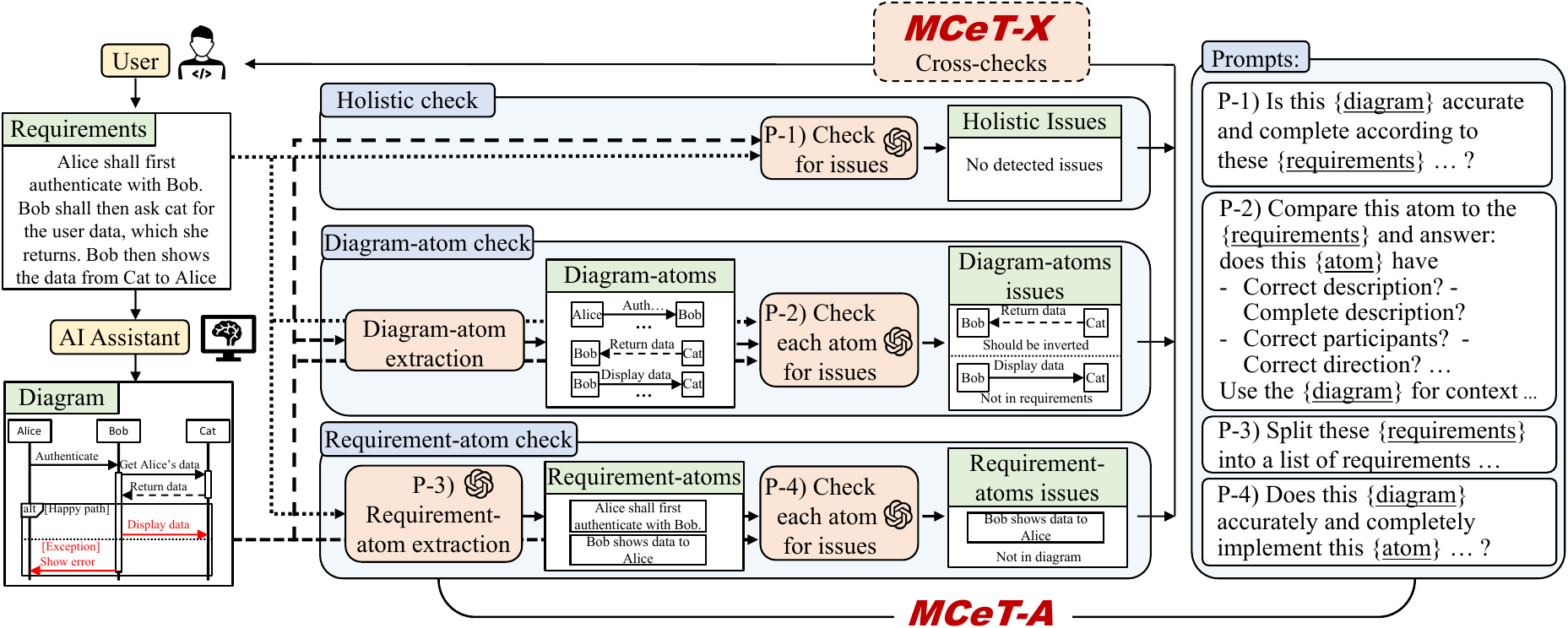}
    \caption{Overview of \tool{}.}
    \vspace{-10pt}
    \label{fig:overview}
\end{figure*}

The overview of \tool{} is shown in Figure~\ref{fig:overview}.
\tool{} takes two inputs, a requirements document and a sequence diagram, possibly generated by an AI assistant.
In our current approach, we process sequence diagrams in PlantUML,
however, the technique can be easily adapted to process any other modeling language.
We first describe how the first stage of \tool, \toolA{}, evaluates the sequence diagram given the requirements by describing each of its three checks,
and then present the authority-based cross-check in the second stage, \toolX.

\subsection{Holistic check}
\label{sec:hol_check}

The holistic check relies on one invocation of the underlying LLM with the prompt P-1, summarized in Figure~\ref{fig:overview}.
The prompt's inputs are the entire diagram, in PlantUML syntax, and the entire requirements text.
The prompt provides few-shot examples~\cite{Dong:Li:Sui:InContextLearning:2024} and asks the LLM to compare the diagram and requirements to each other, use chain-of-thought~\cite{Wei:Wang:Zhou:CoT:NeurIPS:2022} to analyze the differences, and then find accuracy and completeness issues in the diagram.

To improve reliability, we utilize voting~\cite{Chen:Davis:Zou:Voting:NeurIPS:2024,chen2024universal} between several LLM responses for the same prompt.
Specifically, after collecting \code{N} sets of accuracy and completeness issues from \code{N} LLM responses, we do one more LLM invocation, this time asking the LLM to keep issues that appear in \code{N/2} of the responses, and discard the rest.
We use the language capability of LLMs for combining votes as the same issue could be phrased in different ways in each response.

\subsection{Diagram-atoms check}
\label{sec:dia_check}

The first step of this check is to extract atoms from the diagram.
We simply parse the PlantUML file and extract the declaration of each message and its participants to form an atom.
The atom is used as input to the prompt P-2, shown in Figure~\ref{fig:overview}.
The requirements are provided as input so that the accuracy and completeness of the atom are checked against the requirements.
We provide the full diagram as input to the prompt as well to help the LLM understand the context of the atom. For example, the atom \returnRelation{Bob}{Return data}{Cat} is only accurate when it follows the \relation{Bot}{Get Alice's data}{Cat} atom in the diagram.
The prompt instructs the LLM to understand such ordering information and other complex interactions from the diagram.
Finally, the prompt asks questions on the accuracy and completeness of the atom's elements (message, participants, and the direction) and the atom's context.
We also utilize voting in this check where we only keep issues appearing in \code{N/2} of the LLM responses.

\subsection{Requirement-atoms check}
\label{sec:req_check}

This check leverage the reasoning capability of LLMs to split the requirements into requirement-atoms by an LLM invocation using prompt P-3.
For example, the sentence ``Bob shall then ask cat for the user data, which she returns'' is split into ``Bob shall ask cat for the user data'', ``Cat shall return the user data to Bob''.

We then invoke the LLM with prompt P-4 which takes the diagram and requirement atom as inputs, and asks the LLM to check if the diagram implements all elements and interactions mentioned in the atom completely, and to check the accuracy of that implementation.
We also utilize voting for this step.

\subsection{Higher authority cross-check}
\label{sec:cross_check}

As shown in Figure~\ref{fig:motivating_example}, the results from \toolA{} are prone to hallucinations.
For example, \returnRelation{Bob}{Return data}{Cat}: \textit{Should be inverted} is not a real issue.
To mitigate these hallucinations, we propose to perform a self-consistency evaluation across the results of the three checks of \toolA{}.
Typical self-consistency relies on majority voting between several LLM responses~\cite{Chen:Davis:Zou:Voting:NeurIPS:2024},
However, we rely on the observation that each different type of check has its complexities, strengths, and weaknesses~\cite{Cai:Tan:Zhiwen:Gu:FoRAG:SIGKDD:2024}.
We propose to augment voting with a self-consistency evaluation that combines the complexity and strengths of different checks, while attenuating their weaknesses to further reduce hallucinations.

\vspace{-5pt}
\input{sections/tab_prestudy_precision.tex}

Thus, we assess the strengths and weaknesses of each of \toolA's atomic checks; we evaluated the correctness of the issues detected in all three checks on 16 diagrams,
we detail the methodology for this assessment in Section~\ref{sec:evaluation}.
Table~\ref{tab:prestudy_precision} shows the result of this assessment.
The ``Total'' row shows the total number of issues in each of the outputs of the three checks,
the ``True positives'' shows the number of output issues that we deem as real issues,
and the ``Precision'' is the ratio of true positives out of the total detected issues.
The table shows that the precision of the diagram-atom issues is low, only 0.4, compared to that of the requirement-atom issues.

This phenomenon is because a sequence diagram message is always a part of a bigger context, the following example shows a message that only executes under a condition:

\vspace{-2pt}
\messagealt{Report correct active mode}{\lbrack Autopilot is requesting support \rbrack}
\vspace{-2pt}

The task of an LLM is to assess whether this message is accurate, this assessment is context-sensitive.
This is because the reported issue can be related to elements of the atom or the surrounding context, as discussed in Section~\ref{sec:example}.
Thus, \toolA{} reports the following false positive issue for the above message related to its context:
\begin{quoting}
    \vspace{-2pt}
    ... lacks the necessary condition that the autopilot is in a state requesting support, which is essential ...
    \vspace{-2pt}
\end{quoting}

On the other hand, the requirement-atom check does not exhibit this problem,
because the task of the LLM is to assess correctness of the diagram, not the requirements.
This is a simpler task for the LLM than the diagram-atom check as the check is context-free;
the LLM has to only check whether all elements of the requirement atom are implemented accurately in the diagram,
the LLM does not need to extract any context for a requirement-atom to check that the diagram accurately and completely adheres to that atom.

\vspace{-5pt}
\begin{figure}[htbp]
    \centering
    \includegraphics[width=0.7\columnwidth]{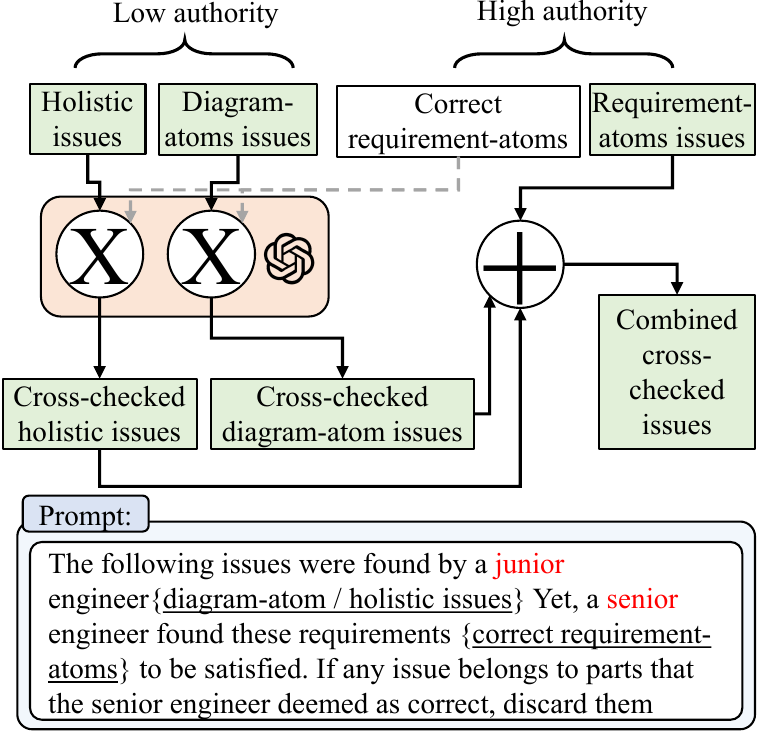}
    \vspace{-5pt}
    \caption{Cross-checking and combining results of different checks.}
    \label{fig:cross_check}
\end{figure}

Indeed, the requirement-atom check detects 6X the number of issues detected by the holistic approach, and 2X the number of issues detected by the diagram-atom approach.
Based on this observation, we propose an approach to leverage the higher precision and recall of the requirement-atom check to automatically filter out false positive issues from the lower precision and recall holistic and diagram-atom checks.
Figure~\ref{fig:cross_check} shows an overview of this approach, which we name \emph{Higher authority cross-check} and adopt in the second stage of our approach~\toolX.

\vspace{-5pt}
\begin{figure}[htbp]
    \centering
    \includegraphics[width=0.75\columnwidth]{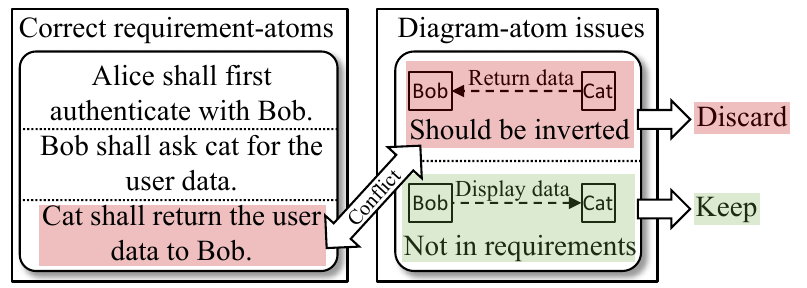}
    \vspace{-5pt}
    \caption{Cross-checking results of the motivating example.}
    \label{fig:cross_check_result}
    \vspace{-10pt}
\end{figure}

First, we identify \textit{Correct Requirement-atoms}, which are requirement-atoms that \toolA{} did not report any issues with.
Then, we perform two independent cross-checks, using two LLM invocations, one for the holistic issues, and the one for the diagram-atom issues.
The cross-check uses the prompt shown in Figure~\ref{fig:cross_check}.
To relay the hierarchy of authority between the issues under check and the correct requirement-atoms, the prompt refers to the issues under check as issues found by a \emph{junior engineer},
while the correct requirement-atoms as requirements found correct by a \emph{senior (higher authority) engineer}.
The LLM is then instructed to find and discard low-authority issues that conflict with correct requirements.

Figure~\ref{fig:cross_check_result} shows the result of applying the cross-check technique on the example from Figure~\ref{fig:motivating_example}.
The correct requirement-atoms are those atoms without issues, i.e., all atoms except for \textit{Bob shows data to Alice}, which is reported as a requirement-atom issue in Figure~\ref{fig:motivating_example}.
All diagram-atom issues reported in Figure~\ref{fig:motivating_example} are cross-checked against the correct requirement-atoms.
The LLM finds that the issue: \returnRelation{Bob}{Return data}{Cat}: \textit{Should be inverted},
is in conflict with the correct requirement-atom \textit{Cat shall return the user data to Bob}.
As the conflicting requirement-atom is deemed correct by a higher authority, we consider the diagram-atom issue as a hallucination and discard it.
We keep diagram atom issues that are not in conflict with any correct requirement-atoms.
For example, the issue \relation{Bob}{Display data}{Cat}: \textit{Not in requirements} is correctly kept.

However, not all diagram-atom issues can be matched with a conflicting requirement-atom issue.
This is because a requirement-atom issue is related to elements of the requirement-atom, which is related to a self-contained part of the diagram, e.g., all  user authentication messages.
However, diagram-atom issues can be related to either elements of the atom, or the context, which could contain several messages, i.e., the previous and next messages, which could span several requirement-atoms.
Thus, not every diagram-atom issue can be matched with one conflicting requirement-atom, cross-checking does not remove these issues that span several requirement-atoms, and retains the unique-perspective of the diagram-atom context issues.

Finally, the kept issues from both the holistic issues and diagram-atom issues, called \textit{Cross-checked holistic issues} and \textit{Cross-checked diagram-atom issues} in Figure~\ref{fig:cross_check},
are combined with the \textit{Requirement-atom issues} to produce the \textit{Combined cross-checked issues} list, the final output of \toolX.

%% file: sections/tab_prestudy_precision.tex
\begin{table}[h]
  \caption{Analysis of issues from 16 diagrams}
  \centering
  \resizebox{0.8\columnwidth}{!}{%
  \begin{tabular}{|l|c|c|c|}
  \hline
  \textbf{\begin{tabular}[c]{@{}c@{}}\tool{} w/\\ \gptmini\end{tabular}} &
    \textbf{\begin{tabular}[c]{@{}c@{}}Holistic\\ issues\end{tabular}} &
    \textbf{\begin{tabular}[c]{@{}c@{}}Diagram-\\ atom issues\end{tabular}} &
    \textbf{\begin{tabular}[c]{@{}c@{}}Requirement-\\ atom issues\end{tabular}} \\ \hline
  \textbf{Total}          & 25   & 75  & 161  \\ \hline
  \textbf{True positives} & 18   & 30  & 142  \\ \hline
  \textbf{Precision}      & 0.72 & 0.4 & 0.88 \\ \hline
  \end{tabular}%
  }
\label{tab:prestudy_precision}
\vspace{-10pt}
\end{table}

%% file: sections/evaluation.tex
\section{Evaluation}
\label{sec:evaluation}

To evaluate \tool{}, we ask the following research questions:

\noindent
\textbf{RQ1 (Precision):}
How effective is \tool{} in detecting real issues versus a baseline holistic approach?

\noindent
\textbf{RQ2 (Human-detected issues):}
How does \tool{} compare with the holistic approach in detecting human-reported issues?

\noindent
\textbf{RQ3 (LLM trade-offs):}
What are the trade-offs of using different reasoning and non-reasoning LLMs in \tool{}?

\subsection{Setup}

\noindent
\textbf{Dataset.}
We evaluate our approach on a set of requirements documents and their scored sequence diagram models which we obtain from Ferrari et al. \cite{Ferrari:Abualhaijal:Arora:REW:2024} and use as our benchmark.
We name this benchmark \textbf{\bench}.
The 28 requirements in this benchmark were collected from various specifications of real industry projects,
including the Lockheed Martin cyber-physical domain requirements~\cite{lm_challenges},
the PURE multi-domains and multi-format requirements~\cite{Ferrari:Spagnolo:Gnesi:RE:2017}, and a dataset of user stories~\cite{Dalpiaz:Fabiano:Sturm:REFSQ:2020}.
Two authors of Ferrari et al. \cite{Ferrari:Abualhaijal:Arora:REW:2024} introduced a set of variants by incrementally introducing requirement smells, such as ambiguity, inconsistency, and incompleteness.
These smells introduce issues in their corresponding generated diagrams, which we aim to detect with \tool.
Our selection of this dataset thus ensures that we evaluate \tool{} on requirements of practical, non-ideal quality.

To generate sequence diagram models, Ferrari et al. \cite{Ferrari:Abualhaijal:Arora:REW:2024} used an LLM (ChatGPT's \code{GPT3.5}) to generate a diagram for each variant.
Each generated diagram was scored on a zero-to-five scale by one of two researchers, who are also experienced software engineers, according to five metrics.
Two of the metrics are \emph{Correctness} and \emph{Completeness}, defined the same as our definition for accuracy and completeness from Section~\ref{sec:example}, respectively.
We focus on these two metrics as \tool{} is designed to find accuracy and completeness issues only.
Whenever the score is not a perfect five, the researchers identified the issues in the diagram that affected the score,
we use these issues as a ground truth of issues found in the diagram by experienced engineers.
More details on the steps taken to ensure the reliability of the ground truth are provided in Ferrari et al.'s paper \cite{Ferrari:Abualhaijal:Arora:REW:2024}.
The total number of variants is 87, we discarded cases where the model generated the wrong type of diagram, e.g., a class diagram, or the researchers did not provide reasons for the score deduction.
The remaining number of variants is 76, which we use to evaluate \tool.



\noindent
\textbf{Subjects.}
We evaluate the outputs of each of the \tool{} checks independently to gain better understanding of how each check contributes to the performance of the tool.
Specifically, we evaluate the \toolA{} output from the holistic check, diagram-atom check, and requirement-atom check.
We also evaluate the \toolX{} outputs for the holistic results and diagram-atom results which are cross-checked against the requirement-atom issues.
Finally, we also evaluate the combined \toolA{} output, and the combined \toolX{} output.
We use the direct holistic check results as a comparison baseline.

\noindent
\textbf{LLMs and \tool{} Configuration.}
We use \gptmini{} within \tool{} to evaluate the diagrams in \bench.
\gptmini{} is selected as it is a lightweight, cost-efficient, and fast LLM with good performance~\cite{OpenAI:GPT4oMini:Online:2024}.
We also repeat the evaluation for a subset of \bench{} using \gpt, \deepv, \deepr.
\gpt~\cite{OpenAI:GPT4o:Online:2024} is a larger, more versatile, and more intelligent version of \gptmini{}.
\deepv~\cite{Liu:Feng:Others:arXiv:2025} is an open-source, large model that employs the Mixture of Experts (MoE) technique that ``activates'' specialized parts of the trained network depending on the input.
\deepr~\cite{DeepSeekAI:Others:DeepSeekR1:arXiv:2025} is an open-source LLM that is trained to employ reasoning, achieves significantly high performance in math, coding, and scientific tasks.
We show through our evaluation that our technique is robust to changes in the underlying LLM, giving good accuracy even on a lighter-weight LLM like \gptmini.

All LLMs are configured with a temperature and top-p of 0.7 and 1, respectively.
The number of votes is set to five.

\subsection{Methodology}
\label{sec:methodology}

To assess the quality of the issues that \tool{} reports, which is required to answer all of our research questions, we rely on the human judgement of two of the authors of this paper.
Both authors are software engineers with 4 and 6 years of experience.
One of the authors participated in teaching a software engineering course for three offerings where he graded sequence diagrams submitted by students.
The other participated in teaching and grading for an algorithms course.

To divide the work of judging the issues among the two authors, we first assessed that both authors have similar judgement such that we can reliably combine their results.
To perform this assessment, we selected the first 20\% of the diagrams (according to the order in the \bench{} paper~\cite{Ferrari:Abualhaijal:Arora:REW:2024}) as a small dataset to assess the similarity on.
Then, we run \tool{} on the selected diagrams, then both authors independently judged the correctness of all detected issues, i.e., each author gives a ``Yes'' or ``No'' rating for the correctness of each issue, independent of the other author's rate.
We then measured the inter-rater reliability using the Cohen's kappa statistic~\cite{Cohen:EPM:1960}, which results in 0.79, indicating a substantial agreement, close to 0.8 which indicates almost perfect agreement.
Afterwards, both authors discussed the rating of issues for which they disagreed.
Both authors settled the disagreements and decided on a common approach of judging any future issues similar to the disputed issues.
The results after settling the disagreements are presented and discussed in Section~\ref{sec:cross_check}.

\input{sections/tab_results_gpt_4o_mini}

\begin{figure}[htbp]
    \centering
    \includegraphics[width=0.9\linewidth]{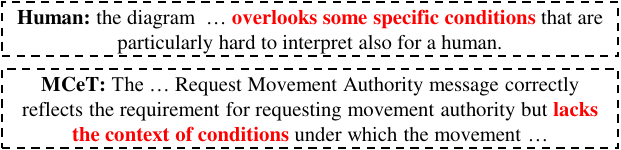}
    \vspace{-5pt}
    \caption{An equivalent pair of a human and an \tool{}-detected issue. The common root issue is missing conditions.}
    \label{fig:human_match}
    \vspace{-5pt}
\end{figure}

To answer RQ1 and RQ2, we split the remaining \tool-detected issues among both authors who proceeded to judge the correctness of each issue.
We record the following metrics:
\textit{\circled{1} Total}: the total number of \tool-detected issues,
\textit{\circled{2} True positives}: the number of correct \tool-detected according to the judging author,
\textit{\circled{3} False positives}: the remaining issues out of ``Total'' that are not ``True positives'',
\textit{\circled{4} Precision}: ratio of ``True positives'' out of the ``Total'' number of issues,
\textit{\circled{4} \bench{} recall}: the number of human-detected issues in \bench{} that are equivalent to at least one \tool-detected issue.
We define equivalent issues as issues that describe the same root cause of the problem in the diagram, even if they have different levels of details.
For example, Figure~\ref{fig:human_match} shows a human-detected issue (top) and an equivalent \tool-detected issue (bottom).
While the human issue is more generic and addresses the missing conditions in the entire diagram, the \tool{} issue focuses on missing conditions in one message, which is a symptom of the same root issue the human reported.
The last metric \textit{\circled{5} New true issues} is the number of ``True positives'' that are not equivalent to any issue reported by the \bench{} authors,
i.e., these are real issues \tool{} found, but \bench{} authors did not find or did not report.

\vspace{-5pt}

To answer RQ3, we re-run \tool{} configured with \gpt{}, \deepv{}, and \deepr{} on 10\% of the dataset (8 diagrams) due to the extensive time and effort required to manually evaluate issues: we evaluated a total of 1524 issues for RQ1 and RQ2, and 347 issues for RQ3.
To ensure we evaluate LLM trade-offs on a wide variety of diagrams, we compute the precision of \tool{} for each case in \bench{} using \gptmini,
we split the 0-1 precision range into 8 equal segments, and randomly select one diagram from each segment.
Thus, we ensure that we repeat our evaluation for different LLMs on a range of diagrams for which \gptmini{}-\tool{} performed both well and poorly.

We compute the metrics used in answering RQ1 and RQ2 with this new configuration.
We also measure
\textit{Average K tokens / diagrams}:
the average number of tokens processed by the LLM invocations within \tool{} per diagram,
and \textit{Average minutes per diagram}: the average execution time of \tool{} in minutes per diagram.
We then identify the trade-off of the precision and the number of human-detected issues reported by \tool{} with the execution time and number of tokens.

%% file: sections/tab_results_gpt_4o_mini.tex
\begin{table*}[htbp]
  \caption{Results of applying \tool{} on the full \bench{} using \gptmini{} as the underlying model}
    \centering
    \resizebox{0.85\textwidth}{!}{%
    \begin{tabular}{|l||cc|cc|c|cc|}
    \hline
    \multicolumn{1}{|c||}{\textbf{Metrics}} &
      \multicolumn{2}{c|}{\textbf{Holistic check (Baseline)}} &
      \multicolumn{2}{c|}{\textbf{Diagram-Atom check}} &
      \multirow{2}{*}{\textbf{\begin{tabular}[c]{@{}c@{}}Req.-Atom check \\ (\toolA{} / \toolX)\end{tabular}}} &
      \multicolumn{2}{c|}{\textbf{Combined checks}} \\ \cline{2-5} \cline{7-8}
    \multicolumn{1}{|c||}{(\textbf{gpt-4o-mini)}} &
      \multicolumn{1}{c|}{\textbf{\toolA}} &
      \textbf{\toolX} &
      \multicolumn{1}{c|}{\textbf{\toolA}} &
      \textbf{\toolX} &
       &
      \multicolumn{1}{c|}{\textbf{\toolA}} &
      \textbf{\toolX} \\ \hline \hline
    \textbf{Total} &
      \multicolumn{1}{c|}{134} &
      69 &
      \multicolumn{1}{c|}{505} &
      201 &
      885 &
      \multicolumn{1}{c|}{1524} &
      1155 \\ \hline
    \textbf{True positives} &
      \multicolumn{1}{c|}{78} &
      43 &
      \multicolumn{1}{c|}{254} &
      131 &
      764 &
      \multicolumn{1}{c|}{1096} &
      938 \\ \hline
    \textbf{False positives} &
      \multicolumn{1}{c|}{56} &
      26 &
      \multicolumn{1}{c|}{251} &
      70 &
      121 &
      \multicolumn{1}{c|}{428} &
      217 \\ \hline
    \textbf{Precision} &
      \multicolumn{1}{c|}{0.58} &
      0.62 &
      \multicolumn{1}{c|}{0.5} &
      0.65 &
      0.86 &
      \multicolumn{1}{c|}{0.72} &
      0.81 \\ \hline
    \textbf{\bench{} recall} &
      \multicolumn{1}{c|}{46 (34.1\%)} &
      27 (20\%) &
      \multicolumn{1}{c|}{43 (31.9\%)} &
      32 (23.7\%) &
      80 (59.3\%) &
      \multicolumn{1}{c|}{92 (68.1\%)} &
      88 (65.2\%) \\ \hline
    \textbf{New true issues} &
      \multicolumn{1}{c|}{27} &
      13 &
      \multicolumn{1}{c|}{138} &
      56 &
      322 &
      \multicolumn{1}{c|}{487} &
      391 \\ \hline
    \end{tabular}%
    }
    \label{tab:all_results}
    \vspace{-10pt}
\end{table*}

%% file: sections/results.tex
\section{Results}
\label{sec:results}

Table~\ref{tab:all_results} shows the result of applying \tool{} on all pairs of sequence diagrams and requirements in \bench{} where \tool{} uses \gptmini{} as the underlying LLM.
The rows of the table are the metrics defined in Section~\ref{sec:methodology}.
The ``Holistic check (Baseline)'' are the issues that are both output directly from the holistic check (\toolA{} column) and from cross-checking (\toolX{} column) against the requirement-atom check results.
The ``Diagram-atom check'' results also include both \toolA{} and \toolX{} issues, while the issues in the ``Req.-Atom check'' column are the same in \toolA{} and \toolX{} since the requirement-atom \toolA{} results are the higher authority used as reference for cross-checking.
The ``Combined checks'' combines all the direct outputs from the holistic, diagram-atom, and requirement-atom checks' results (\toolA{} column),
and combines the cross-checked results from the holistic and diagram-atom checks with the direct results of the requirement-atom check (\toolX{} column).
We use this table to answer both RQ1 and RQ2.

\input{sections/results_rq1.tex}

\input{sections/tab_results_models}

\input{sections/results_rq2.tex}

\input{sections/results_rq3.tex}

%% file: sections/results_rq1.tex
\subsection{RQ1 (Precision)}

As shown in Table~\ref{tab:all_results}, the baseline holistic check's precision (\toolA) is only 0.58, potentially because the holistic check compares the entire diagram with the entire requirements.
When the diagram and requirements are long, the underlying LLM often overlooks details in the prompt which includes both the diagram and the requirements, a phenomenon common for LLMs~\cite{Liu:Lin:Liang:LostMiddle:TACL:2024}, and thus falsely reports that the diagram is not complete.
For example, in the \code{finite\_state\_machine.v0} case~\cite{appendix}, the number of tokens in the requirements and diagram is 953, 2.3X times the average number of tokens of 406 for a requirement-diagram pair in \bench.
The holistic check in this case reports that
\begin{quoting}
    \vspace{-3pt}
    The diagram omits critical details ... including the need to latch a pullup when the pilot is not in control ...
    \vspace{-3pt}
\end{quoting}
Even though the diagram does include the message:

\vspace{-3pt}
\selfmessagealt{Latch autopilot pullup}{\lbrack Autopilot is not in control ... \rbrack}
\vspace{-3pt}

The \toolA{} diagram-atom check also suffers from low precision (0.5) due to underlying LLM failing to extract the context, as observed and reported in the \toolA{} assessment from Section~\ref{sec:cross_check}.
The requirement-atom check achieves a significantly better precision of 0.86, also correlating with \toolA{} assessment results.

Our combined atomic and holistic approach achieves a 0.72 precision which is improved to 0.81 with cross-checking,
indicating that cross-checking eliminates false positive issues.
For example, the following issue is eliminated:
\begin{quoting}
    \vspace{-3pt}
    The message 'Pilot takes \underline{control}' does not specify whether the \underline{pilot is in standby}, which is critical for ...
    \vspace{-15pt}
\end{quoting}
which conflicts with the following requirement-atom that was deemed correctly implemented by the tool:
\begin{quoting}
    \vspace{-3pt}
    The autopilot shall change states from TRANSITION to \underline{STANDBY} when the pilot is in \underline{control} (standby).
    \vspace{-3pt}
\end{quoting}
Since the requirement-atom related to the pilot being in control in the standby mode is deemed as correctly implemented, the related diagram-atom issue is correctly discarded by \toolX.

Cross-checking also removes some real issues, for example, for the following message in \code{g02-uc-cm-req.v0}:
\vspace{-2pt}
\selfmessage{Patient}{Navigate to the website}
\vspace{-2pt}
\toolA{} reports this issue correctly:
\begin{quoting}
    \vspace{-2pt}
    The message ... fails to specify that it is an interaction with the pharmacy website rather than an internal action.
\end{quoting}
i.e., the message should go from \textit{Patient} to \textit{Website} instead of back to \textit{Patient}.
However, the requirement-atom:
\begin{quoting}
\vspace{-2pt}
The patient navigates to the website on his computer
\vspace{-2pt}
\end{quoting}
is deemed as correct, because the LLM finds that the \textit{Navigate to the website} is in the diagram, but fails to detect the issue with the other participant of this message, the website.
Thus, the correctly identified diagram-atom is filtered out.
However, we achieve high precision because cross-checking successfully reduces the number of false positive issues in the combined \toolA{} results from 428 to 217 in the combined \toolX{} results, eliminating 211 false positives,
while only reducing true positives by 158 issues from 1096 to 938.
Thus, \toolA{} is beneficial when higher true positives are desired by the user (e.g., in safety critical systems),
while \toolX{} is useful in exploratory projects, when the user desires less false positives.

We observe that some false positives not filtered out by \toolX{} are caused by vague requirements.
For example, in~\code{1.autopilot.v0}~\cite{Ferrari:Abualhaijal:Arora:REW:2024}, the requirement indicates that the ``autopilot should only be engaged when the pilot selects it'', and also indicates ``autopilot shall engage when pilot selects the autopilot engage switch''.
Both sentences refer to the same requirement and thus are correctly represented once in the diagram.
Yet, the first sentence does not mention ``engage switch'', and thus, \toolX{} assumes that it is not covered by the diagram.
Thus, \toolX{} false positives can help the user spot problems in the requirements.

\noindent
\underline{\textbf{Answer to RQ1}}:
\tool{} achieves a precision of 0.81 by augmenting the holistic check with the atomic and cross-checking approaches, improving over the 0.58 achieved by the baseline holistic-only approach.

%% file: sections/tab_results_models.tex
\begin{table*}[htbp]
    \caption{Trade-offs for using different LLMs within \tool}
    \centering
    \resizebox{\textwidth}{!}{%
    \begin{tabular}{|l||ccc|ccc|ccc|ccc|}
    \hline
    \multicolumn{1}{|c||}{\textbf{Metrics}} &
      \multicolumn{3}{c|}{\textbf{\gptmini}} &
      \multicolumn{3}{c|}{\textbf{\gpt}} &
      \multicolumn{3}{c|}{\textbf{\deepv}} &
      \multicolumn{3}{c|}{\textbf{\deepr}} \\ \cline{2-13}
    \multicolumn{1}{|c||}{\textbf{(10\% of dataset)}} &
      \multicolumn{1}{c|}{\textbf{Baseline}} &
      \multicolumn{1}{c|}{\textbf{\toolA}} & \multicolumn{1}{c|}{\textbf{\toolX}} &
      \multicolumn{1}{c|}{\textbf{Baseline}} &
      \multicolumn{1}{c|}{\textbf{\toolA}} & \multicolumn{1}{c|}{\textbf{\toolX}} &
      \multicolumn{1}{c|}{\textbf{Baseline}} &
      \multicolumn{1}{c|}{\textbf{\toolA}} & \multicolumn{1}{c|}{\textbf{\toolX}} &
      \multicolumn{1}{c|}{\textbf{Baseline}} &
      \multicolumn{1}{c|}{\textbf{\toolA}} & \multicolumn{1}{c|}{\textbf{\toolX}} \\
     \hline \hline
    \textbf{Total} &
      \multicolumn{1}{c|}{14} &
      \multicolumn{1}{c|}{97} &
      75 &
      \multicolumn{1}{c|}{12} &
      \multicolumn{1}{c|}{75} &
      64 &
      \multicolumn{1}{c|}{16} &
      \multicolumn{1}{c|}{70} &
      52 &
      \multicolumn{1}{c|}{24} &
      \multicolumn{1}{c|}{105} &
      65 \\ \hline
    \textbf{True positives} &
      \multicolumn{1}{c|}{5} &
      \multicolumn{1}{c|}{51} &
      43 &
      \multicolumn{1}{c|}{10} &
      \multicolumn{1}{c|}{63} &
      54 &
      \multicolumn{1}{c|}{7} &
      \multicolumn{1}{c|}{45} &
      37 &
      \multicolumn{1}{c|}{17} &
      \multicolumn{1}{c|}{87} &
      57 \\ \hline
      \textbf{False positives} &
      \multicolumn{1}{c|}{9} &
      \multicolumn{1}{c|}{46} &
      32 &
      \multicolumn{1}{c|}{2} &
      \multicolumn{1}{c|}{12} &
      10 &
      \multicolumn{1}{c|}{9} &
      \multicolumn{1}{c|}{25} &
      15 &
      \multicolumn{1}{c|}{7} &
      \multicolumn{1}{c|}{18} &
      8 \\ \hline
    \textbf{Precision} &
      \multicolumn{1}{c|}{0.36} &
      \multicolumn{1}{c|}{0.53} &
      0.57 &
      \multicolumn{1}{c|}{0.83} &
      \multicolumn{1}{c|}{0.84} &
      0.84 &
      \multicolumn{1}{c|}{0.44} &
      \multicolumn{1}{c|}{0.64} &
      0.71 &
      \multicolumn{1}{c|}{0.71} &
      \multicolumn{1}{c|}{0.83} &
      0.88 \\ \hline
    \textbf{\begin{tabular}[l]{@{}l@{}}\bench{}\\ recall\end{tabular}} &
      \multicolumn{1}{c|}{\begin{tabular}[c]{@{}c@{}}3\\ (21.4\%)\end{tabular}} &
      \multicolumn{1}{c|}{\begin{tabular}[c]{@{}c@{}}10\\ (71.4\%)\end{tabular}} &
      \begin{tabular}[c]{@{}c@{}}10\\ (71.4\%)\end{tabular} &
      \multicolumn{1}{c|}{\begin{tabular}[c]{@{}c@{}}4\\ (28.6\%)\end{tabular}} &
      \multicolumn{1}{c|}{\begin{tabular}[c]{@{}c@{}}12\\ (85.7\%)\end{tabular}} &
      \begin{tabular}[c]{@{}c@{}}10\\ (71.4\%)\end{tabular} &
      \multicolumn{1}{c|}{\begin{tabular}[c]{@{}c@{}}5\\ (35.7\%)\end{tabular}} &
      \multicolumn{1}{c|}{\begin{tabular}[c]{@{}c@{}}10\\ (71.4\%)\end{tabular}} &
      \begin{tabular}[c]{@{}c@{}}7\\ (50\%)\end{tabular} &
      \multicolumn{1}{c|}{\begin{tabular}[c]{@{}c@{}}7\\ (50\%)\end{tabular}} &
      \multicolumn{1}{c|}{\begin{tabular}[c]{@{}c@{}}12\\ (85.7\%)\end{tabular}} &
      \begin{tabular}[c]{@{}c@{}}10\\ (71.4\%)\end{tabular} \\ \hline
    \textbf{New true issues} &
      \multicolumn{1}{c|}{2} &
      \multicolumn{1}{c|}{25} &
      19 &
      \multicolumn{1}{c|}{5} &
      \multicolumn{1}{c|}{28} &
      23 &
      \multicolumn{1}{c|}{1} &
      \multicolumn{1}{c|}{16} &
      15 &
      \multicolumn{1}{c|}{7} &
      \multicolumn{1}{c|}{39} &
      25 \\ \hline
      \hline
    \textbf{Avg. $\frac{\text{K tokens}}{\text{diagram}}$} &
      \multicolumn{1}{c|}{12} &
      \multicolumn{1}{c|}{70.8} &
      80.5 &
      \multicolumn{1}{c|}{13.5} &
      \multicolumn{1}{c|}{78} &
      86.7 &
      \multicolumn{1}{c|}{18.1} &
      \multicolumn{1}{c|}{109.2} &
      124.9 &
      \multicolumn{1}{c|}{31.1} &
      \multicolumn{1}{c|}{177.5} &
      236.2 \\ \hline
    \textbf{Avg. $\frac{\text{minutes}}{\text{diagram}}$} &
      \multicolumn{1}{c|}{0.6} &
      \multicolumn{1}{c|}{4.1} &
      4.5 &
      \multicolumn{1}{c|}{0.5} &
      \multicolumn{1}{c|}{4} &
      4.5 &
      \multicolumn{1}{c|}{1.5} &
      \multicolumn{1}{c|}{9.3} &
      10.3 &
      \multicolumn{1}{c|}{8} &
      \multicolumn{1}{c|}{55.4} &
      77.4 \\ \hline
    \end{tabular}%
    }
\label{tab:llm}
\vspace{-10pt}
\end{table*}

%% file: sections/results_rq2.tex
\subsection{RQ2 (Human-detected issues)}

\begin{figure}[htbp]
    \centering
    \includegraphics[width=0.9\linewidth]{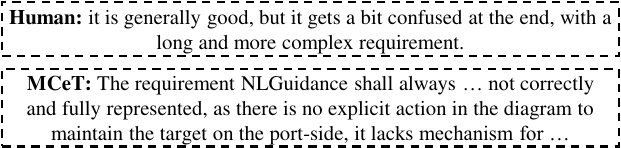}
    \vspace{-5pt}
    \caption{An example of an \bench{} issue and a matching, yet more concrete, issue reported by \tool{}.}
    \label{fig:human_vs_mcet}
    \vspace{-5pt}
\end{figure}

The ``\bench{} recall'' row of Table~\ref{tab:all_results} shows that the combined results of both \toolA{} and \toolX{} significantly outperforms the baseline holistic check, the recall increased from 34.1\% to 68.1\%, and 65.2\% for \toolA{} and \toolX, respectively,
i.e., \tool{} has almost twice the recall of the solitary baseline holistic check.
The atomic check results also add depth compared to the holistic results, as it identify more fine-grained and detailed issues.
Figure~\ref{fig:human_vs_mcet} shows an \bench{} issue and an issue detected by \tool{} from the use case \code{non\_linear\_guidance.v0}~\cite{appendix}.
In this example, the issue reported in \bench{} is too broad and fails to pinpoint the specific problem with the diagram, while \tool-reported issues identify the concrete problematic requirement-atom (NLGuidance), and indicate the concrete problem with the atom (missing actions related to the port-side).

\begin{figure}[htbp]
    \centering
    \includegraphics[width=0.9\linewidth]{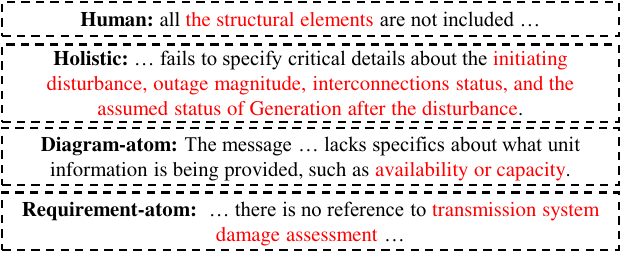}
    \vspace{-5pt}
    \caption{Several issues related to missing structural elements, yet each check indicates a different missing element.}
    \label{fig:multi_issue}
    \vspace{-5pt}
\end{figure}

We observed that the recall of the \toolA{} diagram-atom check is only 31.9\%,
potentially because it is unable to identify issues where entire diagram-atoms are missing.
Since the diagram-atom is missing from the diagram, it cannot be checked for accuracy or completeness.
On the other hand, the requirement-atom check can find if the requirement-atom is not fully implemented, i.e., the diagram misses essential diagram-atoms.
For example, the bottom two issues from Figure~\ref{fig:multi_issue} show a diagram-atom issue and a requirement-atom issue, the requirement-atom issue can find that a whole process is missing from the diagram (the assessment of the system's damage), while the diagram-atom issue focuses on \emph{specifics} of existing diagram-atoms.
Thus, the diagram-atom issues provide a different perspective, despite its lower recall.

\tool{} also identifies new issues that are not reported by the \bench{} authors.
\toolA's diagram-atom and requirement-atom checks identifies 138 and 322 issues overlooked by humans, respectively.
This is nearly 5X and 12X the number of new issues detected by the baseline holistic check.
In the combined \toolA{} results, the total number of new issues discovered rises to 487, which is 18X the issues identified through the direct holistic check.
Even after cross-checking, which filters out some true positive issues, the total number of issues identified in \toolA{} is 391, which is still 14X the issues found through the baseline holistic check.
On average, the direct combined checks is able to identify 6 new issues per diagram, 5 issues after cross-checking.

The total number of true issues for \toolA{} and \toolX{} is 1,094 and 938, which is 14X and 12X the 78 true issues detected by the baseline approach, respectively.
Moreover, combining issues from the three types of checks gives different perspectives to the same problematic part of the diagram.
Figure~\ref{fig:multi_issue} shows a human-reported issue that indicates that all structural elements in the diagram are missing.
The holistic check mentions several missing elements, but not all of them.
The diagram-atom level mentions another missing structural element, and the requirement-atom level mentions yet another missing element.
Combining the three checks together brings the evaluation feedback's comprehensiveness close to that of the human, while also providing concrete details.

\noindent
\underline{\textbf{Answer to RQ2}}:
\tool{} doubles the \bench{} recall, compared to the baseline holistic approach, and identifies 391 new issues compared to human experts, 14x the number of new issues detected by the baseline approach.
\tool's issues provides more concrete, detailed, and multi-perspective insights into the causes of the issues.

%% file: sections/results_rq3.tex
\subsection{RQ3 (LLMs trade-offs)}

Table~\ref{tab:llm} shows the result of applying \tool{} on 8 random diagrams, which is 10\% of the \bench{} using reasoning and non-reasoning LLMs.
The metrics of the table are the same as the metrics of Table~\ref{tab:all_results}, in addition to two metrics:
the ``Avg. K tokens / diagram'' and the ``Avg. minutes / diagram''.
Each three consecutive columns are the results for one LLM, e.g., columns 1-3 are the results when \tool{} uses \gptmini{} as the underlying LLM.
The columns correspond to the holistic check output, combined direct output, and cross-checked combined output, respectively.

The precision row in Table~\ref{tab:llm} shows that the atomic approach \emph{consistently improves} upon the holistic only approach;
the lighter-weight LLMs benefit the most: \gptmini's and \deepv's precision improved from 0.36 to 0.57 and from 0.44 to 0.71, respectively.
Even the more sophisticated LLM \gpt's precision improved from 0.83 to 0.84.
The reasoning LLM \deepr's precision also improved from 0.71 to 0.88.

Our approach improves the recall of \bench{} issues from the holistic baseline for the light-weight \gptmini{} LLM,
and significantly improves the recall of the more-sophisticated \gpt{} and the reasoning \deepr{} LLMs; the recall is improved from 28.6\% to 71.4\% and from 50\% to 71.4\%, respectively.
Note that \deepr{} starts with a higher holistic recall (50\%) than the other LLMs.
This is because the LLM's reasoning attempts a fine-grained evaluation, similar to the \tool{} approach, for example:
\begin{quoting}
    \vspace{-5pt}
    ... First, I'll go through each requirement and check how it's represented in the diagram ...
    \vspace{-5pt}
\end{quoting}
Here, the LLM does a less structured version of our requirement-atom checking, however, it does not focus on each and every atom, it does not combine the analysis from both requirement-atom and diagram-atom perspectives, and does not have information about higher authority results to perform our cross-check.
Thus, even the reasoning LLM benefit from \tool{}'s approach, both in precision and recall.
Only \deepv{} has a drop in recall (from 71.4\% to 50\%), however, \deepv{} cross-checked results still outperform the baseline holistic precision and recall.

\deepr{} and \gpt{} achieve the best precision and recall, but cost more tokens and time per diagram than the lighter-weight \gptmini{}.
The best overall precision is achieved by \deepr{} cross-checked result (0.88 precision) but consumes 2.7X the tokens and 17.2X the time that the runner-up \gpt{} (0.84 precision) takes.
\deepr{} \toolX{} also consumes 2.9X the tokens 17.2X the time than \gptmini{} while achieving the same recall.
Thus, \tool{} enables non-reasoning LLMs to achieve comparable performance to reasoning LLMs at a fraction of the cost.

Moreover, the light-weight \gptmini's \toolX{} recall (71.4\%) is even higher than \deepr's holistic recall (50\%),
while \gpt's \toolX{} precision and recall (0.84, 71.4\%) are both higher than \deepr's holistic precision and recall (0.71, 50\%).
Both  \gptmini{} and \gpt{} \toolX{} require only 4.5 minutes, 0.56X the time that \deepr{} requires for the holistic evaluation.
Thus, \toolX{} meets or surpasses a reasoning-LLM based baseline holistic-only approach with a fraction of the cost.

\noindent
\underline{\textbf{Answer to RQ3}}:
\deepr{}-\tool{} achieves the best performance (precision and recall), but requires 2.7X the tokens and 17.2X the time than \gpt{}-\tool{} which achieves comparable performance.
\tool{} moreover enables \gptmini{} and \gpt{} to surpass \deepr{} baseline holistic performance with 0.56X execution time.

%% file: sections/threats.tex
\section{Threats to Validity}
\label{sec:threats}

\noindent
\textbf{Internal Validity.}
The subjective judgments from the two authors influences the evaluation of \tool.
To mitigate the potential bias, we computed Cohen's Kappa coefficient~\cite{Cohen:EPM:1960} between the two authors, which indicated substantial agreement.
Both authors discussed and standardized the criteria for evaluating issues.
Moreover, results for RQ3 may not generalize for the entire dataset.
To mitigate this issue, we selected 8 diagrams that span the entire precision spectrum in RQ1, i.e., we selected diagrams for which \tool{} performed both well and poorly on the full dataset,
to ensure that the RQ3 results are more likely to be reflective of the entire dataset.

\noindent
\textbf{External Validity.}
LLMs are probabilistic in nature, it is possible to get different issues each time \tool{} is run on a use case.
To address this consistency issue, we implemented a voting mechanism, combining the majority response from the answers.
Finally, our approach may not generalize to other types of behavioral models, we aim to expand our study into other types of behavioral models as part of our future work.

%% file: sections/related.tex
\section{Related Work}
\label{sec:related}

\noindent
\textbf{LLM-as-a-Judge.}
With the rise of LLMs, their use in replacing labor-intensive evaluation has attracted increasing interest \cite{Chiang:Lee:ACL:2023, Want:Liang:Others:Zhou:2023, gu2024survey,li2024llms}, with most approaches distinguishable by LLM's role in the evaluation process.
A common category of methods extracts generation probabilities from the LLM as evaluation scores \cite{zhang2020bertscore,Fu:Ng:Jiang:Liu:NAACL:2024,jia2023zero}. For example, GPT-Score \cite{Fu:Ng:Jiang:Liu:NAACL:2024} computes the conditional probability of the solution given the task description as the score. Another class of approaches prompts LLMs to directly output evaluation scores using a predefined grading scheme \cite{lin2023llm,liu2023g}. This direct scoring strategy also enables the use of multiple LLMs in a multi-agent evaluation system \cite{chan2024chateval,liang2024abseval}. In the context of software engineering, Wang et al. \cite{wang2025can} systematically evaluated the performance of LLM-as-a-Judge on various coding tasks.

A related line of work involves using LLMs for self-reflection to enhance generation quality \cite{renze2024self,pan2023automatically}. Methods such as Reflexion \cite{shinn2023reflexion} and Self-Refine \cite{madaan2023self} aim to identify errors in current LLM outputs and provide feedback for subsequent generations, either to avoid or correct these mistakes.

\tool{} leverages the LLM-as-a-Judge paradigm to detect issues in a sequence diagram by comparing it with the original textual requirements' description.
Moreover, \tool{} uses a finer-grained evaluations approach by breaking down the evaluation into different checks. Since each check focuses on different aspects of model quality, we further propose a \emph{corss-check} strategy that combines strength of different checks to mitigate hallucinations from LLMs.

\noindent
\textbf{Automated model evaluation.}
Many approaches target automatically evaluating the quality of models,
typically in an educational setting where an instructor-created \emph{reference model} is available \cite{bian2020automated}.
When a reference model is utilized, automated evaluation methods generally compare candidate models against this reference, identifying differences through rule-based or graph-matching approaches tailored specifically for UML model comparison \cite{tselonis2005diagram}. Such comparison tool usually built around popular modeling frameworks such as the Eclipse Modeling Framework (EMF) \cite{emf_diffmerge, EMFCompare} and TouchCore \cite{bian2019automated, hosseinibaghdadabadi2023automated}. Based on identified mismatches, errors can be systematically extracted \cite{singh2022detecting}. More recently, embedding-based techniques have complemented traditional graph-matching by capturing semantic similarities between model elements, further improving evaluation performance \cite{chen2024embedding, triandini2019sequence}.

However, a critical limitation of reference-based evaluation approaches is that they fail to consider alternative yet valid models. Furthermore, in many real-world scenarios, a suitable reference model may not even be available. Consequently, alternative methods aim to identify common modeling mistakes independently of reference solutions, typically employing rule-based heuristics \cite{hasker2011umlint} or machine learning classifiers \cite{boubekeur2020automatic, stikkolorum2019towards}. For instance, Boubekeur et al. \cite{boubekeur2020automatic} extracted heuristic rules based on a provided marking scheme to train classifiers predicting the quality of domain models.

Compared to existing methods, by utilizing LLMs, \tool{} can directly compare candidate models against the text requirement descriptions, allowing the extraction of modeling errors without relying on explicit rules or training data.
Although this paper primarily focuses on evaluating the sound foundation of models against the requirement, \tool{} can be effectively combined with complementary approaches to identify and correct broader issues related to modeling practices.



\noindent
\textbf{LLM and MDE.}
In the context of model-driven engineering (MDE), the application of LLMs has become increasingly popular in modeling tasks, including model generation, completion, transformation, and the development of tools for domain-specific languages (DSLs) \cite{di2025use}.
For model generation, LLMs are used to convert textual descriptions into complete model artifacts, such as domain models \cite{chen2023automated,arulmohan2023extracting}, sequence diagrams \cite{Ferrari:Abualhaijal:Arora:REW:2024,jahan2024automated}, and goal models \cite{chen2023use}.
In some cases, LLMs assist human modelers during the modeling process, e.g., in model completion \cite{chaaben2023towards,apvrille2024system} and in model transformation \cite{buchmann2024prompting}.
Furthermore, LLMs can be utilized to translate natural language into DSL queries, such as VQL \cite{lopez2024text2vql} and OCL \cite{abukhalaf2023codex}.

Despite the promising capabilities of LLMs to generate model diagrams directly from textual input, the generated results often suffer from hallucinations, leading to misalignment with the original descriptions \cite{chen2023automated}. In this paper, we propose \tool{}, an approach for automatically detecting such misalignments as issues. \tool{} can serve as an automated evaluation mechanism for LLM-generated models or provide feedback for manual or automated correction of these issues.

%% file: sections/conclusions.tex
\section{Conclusions}
\label{sec:conclusions}

This paper presents a behavioral model evaluation approach, implemented in the tool \tool{},
which evaluates a sequence diagram against the requirements and identifies and reports potential issues.
\tool{} leverages the language reasoning capabilities of LLMs and enhances them with fine-grained, multi-perspective, and self-consistency evaluation.
To the best of our knowledge, \tool{} is the first approach to perform fully automated evaluation of a behavioral model against its requirements.
We evaluated \tool{} on real requirements and their corresponding sequence diagrams and showed that it achieves high precision and human-like issue detection capabilities.
We plan to investigate preventing the cross-checking approach from eliminating real issues, and generalizing our approach to more types of models in our future work.